\documentclass[[10pt,a4paper,english]{article}
\pdfoutput=1
\usepackage[utf8]{inputenc}
\usepackage[T1]{fontenc}
\let\tldocfrench=1 
\usepackage{csquotes}
\usepackage{babel}
\usepackage{xspace}

\usepackage{tex-live}
\usepackage{amsmath}
\usepackage{amsfonts}
\usepackage{amssymb}
\usepackage{tabularx}
\usepackage{array}
\usepackage{cases}
\usepackage{graphicx}
\usepackage{lmodern}
\usepackage{makeidx}
\usepackage{xkeyval}

\ifpdf
\usepackage{hyperref}
\usepackage[all]{hypcap}
\fi
\begin{document}
\hypersetup{pdfauthor={some author},pdftitle={eye-catching title}}
\author{Noureddine HADJI $^{\href{https://orcid.org/0000-0001-5876-8002}{\includegraphics[scale=.2]{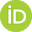}},}$ $^{{\href{http://arxiv.org/a/hadji_n_2}{\color{blue}\textrm {arXiv-id}}}}$} 
\title{{\bf Diffraction from inhomogeneous material systems.}}
\maketitle
Department of Physics and Astronomy, University of Glasgow, Glasgow G12 8QQ, Scotland. UK. \\
\textit{Present address:} \noindent D\'{e}partement de Physique, Universit\'{e} Badji Mokhtar, Annaba, BP 12 Annaba 23000, Alg\'{e}rie.\\
\textit{Email:} noureddine.hadji@univ-annaba.dz \\
\\
The static diffraction intensity distribution from large material system conceived as perfectly homogeneous system made inhomogeneous, though substitution of groups of atoms, small particles, by other groups of atoms, is explicitly expressed in terms of all possible partial contributions subsequent to the particle exchange operation. This gives the diffracted intensity as that from the homogeneous system modified by additional part and, thus, yields result that may prove to be quite useful in comparative studies of materials with the same sort of atomic network. The additional
part embodies the effects of the presence of the structural disturbances and consists in two main components. One of these represents the inhomogeneity coupling with the system in its homogeneous state and, the other, represents the inhomogeneity self-coupling, the coupling of the inhomogeneities with one another.  Two particular cases of deviation from homogeneity are considered in more details: the case of non-interacting inhomogeneities and that of interacting inhomogeneities. The effects on the small angle scattering (SAS) due to possible structural differences (chemical nature, atomic distribution and dimensions) between the deficient and substitute particles are given. The SAS distribution does not obey the very attractive Guinier law in some cases of non-interacting inhomogeneities.
\tableofcontents
\listoffigures
%
\section{Introduction}
The studies of real materials through diffraction are of undeniably great value and allow obtaining structural information on the atomic distribution as well as on the distribution, size and shape of inhomogeneities present within materials. Substantial efforts have been made in the treatment of the problems relative to diffraction from scattering systems in order to determine the materials structural parameters such as the atomic distribution, the sizes and shapes of small particles etc. resulting in a very rich
diffraction-related literature, for instance naming just a few \cite{Friedrich Knipping and Laue 1913}-\cite{Grawert and Svergun 2020}. Here, the total scattered intensity distribution for a material system whose homogeneity has been disturbed through substitution of groups of atoms by other groups of atoms, hereafter termed, respectively, deficient particles and substitute particles, of different chemical nature and different atomic distribution is obtained and partially studied. Effectively, the expression of the total elastic SID, also referred to as total SID, is obtained as an explicit function of the scattering components originating (i) from the different structural sub-systems, namely the atomic, deficient-particles and substitute-particles networks, that make up the material system and (ii) from the different couplings between these sub-systems. The calculations were carried out for real atomic scattering factors and within the frame of the kinematic approach to diffraction making use of no special kind-of-material-related structural assumption. Therefore, the obtained results should hold for the wide range of material systems for which the atomic scattering factors can be assumed to be real. The derived expression for the total SID conveniently gives the modifications that are introduced on the scattering distribution from a material system in its homogeneous state by the presence of structural disturbances, or, which is equivalent, of inhomogeneities. Therefore this expression may prove to be quite useful for obtaining the effects on the diffracted intensity due to particular kinds of disturbances or even for obtaining the diffracted intensity relative to situations such as those in which the material systems consist only of groups of atoms of different sorts, that is of groups of atoms which are not embedded in an otherwise homogeneous material system. An example of these can be the particular case of particles possessing a center of symmetry, with the note that the idea of `a center of symmetry' was first used by Guinier, \cite{Guinier (1939)}, in scattering studies connected with ensembles of particles (groups of atoms), see also \cite{Guinier and Fournet (1955)}. As a contribution to this case a full section, the fifth, is devoted to the expression for the SID from inhomogeneities that possess a center of symmetry. The implications for the small angle scattering (SAS) of possible structural differences in chemical nature and atomic distribution between the deficient and substitute particles have been given a particular attention here. This has led to a shape of the small angle SID that, in some cases of structural disturbances, does not obey the very attractive ``Gaussian'' Guinier law, see \cite{Guinier (1939), Guinier and Fournet (1955), Guinier (1963)} for the detailed description of this law. Consequently, the techniques of analysis which can be used to extract the so-called radius of gyration, e.g. see \cite{Guinier (1939), Guinier and Fournet (1955), Guinier (1963), Hadji and Craven (1988)} for these techniques, from real substances are not valid in some cases and, thus, new techniques will be welcome for these cases.\\
The main expression for the diffracted intensity obtained here applies to a wide range of material systems, which makes it eligible for use to immediately infer diffraction intensity expressions relevant to these systems. It is thus that, as examples, (a) a number of, already existing and non existing, SAS-related results for the diffracted intensity are deduced from the main expression, given below by (\ref{4}) and therefore by (\ref{5}), and presented under a table sort of scheme and (b) a full section is dedicated to the case of identical interacting inhomogeneities.\\
No particular study of the modifications that are introduced on the medium and large angle scattering is considered here.\\
Only time-independent, static, diffraction is considered here.
\section{General situation: the expression of the total static elastic SID}
In this first section, the broad situation generated by inhomogeneities present within real materials is described and the different resulting components of the expression of the total elastic scattering intensity separated.\\
The usual expression for the scattered intensity distribution from a material system is (e.g. \cite{Guinier and Fournet (1955)}, \cite{Guinier (1963)}, \cite{Warren (1990)}, \cite{Cowley (1992-1993)})
\begin{equation}
I(\vec{S})=\sum_{P=1}^{Nr}\sum_{Q=1}^{Nr}f_{P}(\vec{S})f_{Q}(\vec{S})\cos (%
\vec{S}.\vec{r}_{PQ})=\sum_{P=1}^{Nr}\sum_{Q=1}^{Nr}J_{PQ}\text{,}  \label{1}
\end{equation}
where the summations are over all atoms of the material ($Nr$ in total). $f_{P(Q)}$ is the scattering factor of atom $P$($Q$); $\vec{r}_{PQ}$ is the vector that gives the relative positions of atoms $P$ and $Q$; $\vec{S} =(4\pi /\lambda $)$\sin \theta $.$\vec{u}$ ($\lambda =$ wavelength, $2\theta $ is the scattering angle and $\vec{u}$ the unit vector that defines the direction of the scattering vector $\vec{S}$). Defining, on the one hand, that: 1) $N$ is the total number of atoms in the homogeneous material before the substitutions take place; 2) $n$ is the total number of the deficient particles. And, assuming, on the other hand, (a) that the $i^{th}$ deficient and the $i^{th}$ substitute particles are made up of, respectively, $M_{2i}$ and 
$M_{1i}$ atoms, and (b) for simplicity reasons, that the total number of the deficient particles is the same as that of the substitute particles allows writing the total number, $Nr$, of atoms in the inhomogeneous material as
\begin{equation}
Nr=N+\sum_{i=1}^{n}M_{1i}-\sum_{i=1}^{n}M_{2i} \text{.}  \label{2}
\end{equation}
And noticing that
\[
\cos (\vec{S}.\vec{r}_{PQ})=\cos (\vec{S}.(-\vec{r}_{QP}))=\cos (\vec{S}.%
\vec{r}_{QP}) 
\]
allows writing
\begin{equation}
J_{PQ}=J_{QP}.  \label{3}
\end{equation}
This means that $J_{PQ}$ is an invariant with respect to the permutation of the labels $P$ and $Q$ and that we can write (\ref{1}) as (M.13), see proof at the end of the present section, or as
\begin{eqnarray}
I(\vec{S})
&=&\sum_{P=1}^{N}\sum_{Q=1}^{N}J_{PQ}+2\sum_{P=1}^{N}\left\lbrace \sum_{i=1}^{n}\left[ %
\sum_{q_{i}=1}^{M_{1i}}J_{Pq_{i}}-\sum_{q_{i}^{\prime
}=1}^{M_{2i}}J_{Pq_{i}^{\prime }}\right] \right\rbrace   \nonumber \\
&&+\sum_{i=1}^{n}\sum_{j=1}^{n}\left\lbrace \sum_{p_{i}^{\prime
}=1}^{M_{2i}}\sum_{q_{i}^{\prime }=1}^{M_{2i}}J_{p_{i}^{\prime
}q_{j}^{\prime }}-2\sum_{p_{i}^{\prime
}=1}^{M_{2i}}\sum_{q_{j}=1}^{M_{1j}}J_{p_{i}^{\prime }q_{j}}
+\sum_{p_{i}=1}^{M_{1i}}\sum_{q_{j}=1}^{M_{1j}}J_{p_{i}q_{j}}\right\rbrace \text{,}
\label{4}
\end{eqnarray}
where, now, with the new notations, $P$ and $Q$ are the labels of the atoms of the material in its homogeneous state, if no inhomogeneity were present, $p_{i(j)}$ and $q_{i(j)}$ are the labels of the atoms of the $i(j)^{th}$ substitute particle, $p_{i(j)}^{\prime }$ and $q_{i(j)}^{\prime }$ are the labels of the atoms of the $i(j)^{th}$ deficient particle, $i$ and $j$ being the labels of both types of particles.\\
Also, noticing that 
\begin{eqnarray*}
\sum_{k_{u}=1}^{N_{u}}\sum_{m_{v}=1}^{N_{v}}J_{k_{u}m_{v}}
&=&\sum_{k_{u}=1}^{N_{u}}\sum_{m_{v}=1}^{N_{v}}f_{k_{u}}f_{m_{v}}\cos \vec{S}%
.\vec{r}_{k_{u}m_{v}} \\
&=&\left[ \sum_{m_{v}=1}^{N_{v}}f_{m_{v}}\exp (-i\vec{S}.\vec{r}%
_{m_{v}})\right] \left[ \sum_{k_{u}=1}^{N_{u}}f_{k_{u}}\exp (i\vec{S}.\vec{r}%
_{k_{u}})\right] \\
&=&\digamma _{v}\digamma _{u}^{*}
\end{eqnarray*}
is the product of the structure factor $\digamma _{v}$ {\large (}e.g. of the $v^{th}$ deficient (or substitute) particle (with $N_{v}=M_{1(2)}$) or of the system in its homogeneous state (with $N_{v}=N$){\large )} and its complex conjugate $\digamma _{u}^{*}$ {\large [}e.g. of the $u^{th}$ deficient (or substitute) particle (with $N_{u}=M_{1(2)}$) or of the system in its homogeneous state (with $N_{u}=N$){\large ]} allows rewriting (\ref{4}) as
\begin{equation}
I(\vec{S}) =\digamma _{H}\digamma _{H}^{*}+2\sum_{i=1}^{n}\left[ \digamma
_{i}-\digamma _{i}^{\prime }\right] \digamma
_{H}^{*}+\sum_{i=1}^{n}\sum_{j=1}^{n}\left\lbrace \left[ \digamma _{j}^{\prime }-\digamma
_{j}]\digamma _{i}^{\prime *} 
+\digamma _{j}[\digamma _{i}^{*}-\digamma _{i}^{\prime *}\right] \right\rbrace \text{,}
\label{5}
\end{equation}
where $i$ and $j$ are relative to the different substitute and deficient particles.\\
Moreover, we may write (\ref{4}), and therefore (\ref{5}), as:
\begin{equation}
I(\vec{S})=I_{HH}(\vec{S})+I_{ad}(\vec{S})=I_{HH}(\vec{S})+I_{HD}(\vec{S}%
)+I_{DD}(\vec{S})\text{,}  \label{6}
\end{equation}
where $I_{HH}(\vec{S})$, corresponds to the first terms of equations (\ref{4}) (and (\ref{5})) and represents the scattered intensity in absence of inhomogeneities; $I_{ad}(\vec{S})$ corresponds to the remaining terms of (\ref{4}), and therefore of (\ref{5}), and represents the resulting effect of the generation of inhomogeneities in the material system. The total scattered intensity, as expressed by equations (\ref{4}) and (\ref{5}), may, therefore, be thought of as the scattering, $I_{HH}(\vec{S})$, from the material in its homogeneous state taken as reference modified by an additional term, $I_{ad}(\vec{S})$. This term represents the deviation from homogeneity of the real material. $I_{ad}(\vec{S})$, made up of two terms $%
I_{HD}(\vec{S})$ and $I_{DD}(\vec{S})$, see (\ref{6}), may be expressed as:
\[
I_{ad}(\vec{S})=I_{HD}(\vec{S})+I_{DD}(\vec{S})\text{,}
\]
where
\begin{eqnarray}
I_{HD}(\vec{S}) &=&I_{HD1}(\vec{S})+I_{HD2}(\vec{S})  \nonumber \\
&=&2\sum_{P=1}^{N}\left\lbrace \sum_{i=1}^{n}\left[ \sum_{q_{i}=1}^{M_{1i}}f_{P}f_{q_{i}}\cos
(\vec{S}.\vec{r}_{Pq_{i}})-\sum_{q_{i}^{\prime
}=1}^{M_{2i}}f_{P}f_{q_{i}^{\prime }}\cos (\vec{S}.\vec{r}_{Pq_{i}^{\prime
}})\right] \right\rbrace   \label{7}
\end{eqnarray}
is the contribution from the pairing of atoms ($P$) of the system in its homogeneous state with atoms $q_{i}$ and $q_{i}^{\prime }$ of the ensemble of the local deviations ($i$). Each of these deviations is represented by one deficient and one substitute particles. It may also be thought of as representing the coupling between the material in its homogeneous state and the ensemble of the local deviations present in the real material.\\ $I_{DD}(\vec{S}$), given by:
\begin{eqnarray}
I_{DD}(\vec{S}) &=&\sum_{i=1}^{n}\sum_{j=1}^{n}\left\lbrace \sum_{p_{i}^{\prime
}=1}^{M_{2i}}\sum_{q_{j}^{\prime }=1}^{M_{2j}}f_{p_{i}^{\prime
}}f_{q_{j}^{\prime }}\cos (\vec{S}.\vec{r}_{p_{i}^{\prime }q_{j}^{\prime }})\right.
\nonumber \\
&& \left.-2\sum_{p_{i}^{\prime }=1}^{M_{2i}}\sum_{q_{j}=1}^{M_{1j}}f_{p_{i}^{\prime
}}f_{q_{j}}\cos (\vec{S}.\vec{r}_{p_{i}^{\prime
}q_{j}})+\sum_{p_{i}=1}^{M_{1i}}\sum_{q_{j}=1}^{M_{1j}}f_{p_{i}}f_{q_{j}}%
\cos (\vec{S}.\vec{r}_{p_{i}q_{j}})\right\rbrace \text{,}  \label{8}
\end{eqnarray}
does not involve any atom of the ``homogeneous system'' and is associated with the local deviations acting as if they were totally isolated from the host material. These considerations allow saying that the total scattering from inhomogeneous materials may not be considered as a simple sum of the scattering from the ideal network [that which is associated with the material in its homogeneous state ($I_{HH}(\vec{S})$)] and the scattering from the inhomogeneities in absence of the host material: the insertion of perturbations results in the appearance of extra terms that are inherent in the total scattered intensity. These extra terms are: 
\begin{itemize}
\item $I_{HD}(\vec{S})$, the term that represents the coupling between the homogeneous system and the
deviations,
\item $I_{DD1}(\vec{S})$, this is due to the deficient particles acting as is they were isolated from the rest of the material and \item $I_{DD2}(\vec{S})$ which represents the coupling between the deficient and the substitute particles.
\end{itemize}  
The formulation of the SID from a real material system as given by (\ref{6}) may prove to be particularly useful in comparative studies of amorphous materials that have identical atomic networks. The subtraction, see \cite{Hadji et al (1987)}, of $I_{1}(\vec{S})$, the scattering profile obtained from an amorphous specimen 1, from $I_{2}(\vec{S})$, obtained from a different area within specimen 1 or from a specimen 2 that has the same network and the same thickness as specimen 1, will reduce the contributions that are equal, e.g. those from the homogeneous networks, to nothing leaving a difference that is due to the existence of dissimilarities in the distributions of inhomogeneities present within the specimens. \\
(\ref{6}) is the general expression of the static elastic scattering intensity for an \emph{arbitrary} inhomogeneous material system; and we believe that it could be helpful to keep it within arm's reach as it can be quite useful when the interest is in rapidly getting the particular intensity expression corresponding to the particular inhomogeneous material system in hand. This is for the reason that this (\ref{6}), constituting the central result of the whole work presented here, can be seen to represent at least a large part of the 'kinematic approach to diffraction'-related results already existing in the literature, for instance see \cite{Guinier (1939)}, \cite{Guinier and Fournet (1955)}), \cite{Guinier (1963)}, \cite{Warren (1990)}-\cite{Grawert and Svergun 2020}, as well as, possibly, results for quite a number of not yet studied heterogeneous material systems. It is thus that the remaining part of this work is dedicated to applying the third component, given by (\ref{8}), of this (\ref{6}) to several not complicated particulate systems. In the immediately following section, examples of diffraction intensities for particular such systems cases are inferred using (\ref{8}). The section just after the next one, Sec. \ref{Examples2}, deals with the so called small angle scattering and the one after, Sec. \ref{Examples3}, deals with the case of the diffraction from identical centrosymmetric interacting local deviations.\\
\rule[0mm]{\linewidth}{3.pt}
\paragraph{{\large A useful intensity-related result:}}
Here we are concerned with the splitting of a rather complex double sum of the type of
\begin{equation}
A=\sum_{P=1}^{N_{r}}\sum_{Q=1}^{N_{r}}J_{PQ}  \tag{M.1}
\end{equation}
into its different components to have a much easier to deal with form, the form given by (M.13) below. In (M.1) the labels $P$ and $Q$ take the same values and refer to the atoms within the \emph{inhomogeneous} specimen, the total number of these being $N_{r}$ and given by:
\begin{equation}
N_{r}=N-\sum_{i=1}^{n}M_{2i}+\sum_{i=1}^{n}M_{1i}\text{,}  \tag{M.2}
\end{equation}
where $ N $ is the total number of atoms within the specimen in its \emph{homogeneous} state, $ n $ is the total number of inhomogeneities within the specimen and the $i^{th}$ inhomogeneity consists of $M_{2i}$ deficient atoms and $M_{1i}$ replacement atoms. But in facts, the specimen consists of a matrix and inclusion groups of atoms; the specimen matrix is made up of $N-\sum_{i=1}^{n}M_{2i}$ atoms and the $i^{th}$ inclusion group of atoms is made up of $M_{1i}$ atoms.
\\
\textbf{{\large Proof:}} Writing symbolically $Nr=N-M$, where the ``symbol'' $M$ is given by: 
\begin{equation}
M=\sum_{i=1}^{n}M_{2i}-\sum_{i=1}^{n}M_{1i}\text{,}  \tag{M.3}
\end{equation}
allows rewriting (M.1) concisely as: 
\begin{equation}
A=\sum_{P=1}^{N-M}\sum_{Q=1}^{N-M}J_{PQ}\text{,}  \tag{M.4}
\end{equation}
and then as 
\begin{equation}
A=\sum_{P=1}^{N-M}\left( \sum_{Q=1}^{N}J_{PQ}-\sum_{Q=(N-(M-1))}^{N}J_{PQ}\right) \text{,}
\tag{M.6}
\end{equation}
where the sum $\sum_{Q=(N-(M-1))}^{N}$ is over the atoms whose number is represented symbolically by $M$, see (M.3), and therefore can be rewritten to reflect this fact, for instance making use of the the following substitutions $Q\longrightarrow Q_{1}=Q-(N-M)=1$ and $N\longrightarrow M=N-(N-M)$, as $\sum_{Q_{1}=1}^{M}$. Thus (M.6) can be written as:
\begin{equation}
A=\sum_{P=1}^{N-M}\sum_{Q=1}^{N}J_{PQ}-\sum_{P=1}^{N-M}\sum_{Q_{1}=1}^{M}J_{PQ_{1}}\text{,}  \tag{M.7}
\end{equation}
and can be split further to be:
\begin{equation*}
A=\left\lbrace \sum_{P=1}^{N}\sum_{Q=1}^{N}J_{PQ} - \sum_{P=N-(M-1)}^{N}\sum_{Q=1}^{N} J_{PQ} \right\rbrace \
-\left\lbrace \sum_{P=1}^{N}\sum_{Q_{1}=1}^{M}J_{PQ_{1}}-\sum_{P=N-(M-1)}^{N}\sum_{Q_{1}=1}^{M}J_{PQ_{1}} \right\rbrace \tag{M.8}
\end{equation*}
where the sum $\sum_{P=(N-(M-1))}^{N}$ is also over the atoms whose number is represented symbolically by $M$ and can, also, be rewritten to reflect this fact, e.g. as $\sum_{P_{1}=1}^{M}$. This leads to:
\begin{equation*}
A=\left\lbrace \sum_{P=1}^{N}\sum_{Q=1}^{N}J_{PQ} - \sum_{P_{1}=1}^{M}\sum_{Q=1}^{N} J_{P_{1}Q} \right\rbrace
-\left\lbrace \sum_{P=1}^{N}\sum_{Q_{1}=1}^{M}J_{PQ_{1}}-\sum_{P_{1}=1}^{M}\sum_{Q_{1}=1}^{M}J_{P_{1}Q_{1}} \right\rbrace \tag{M.9}
\end{equation*}
where therefore $Q_{1}$ and $P_{1}$ take the same values and are both labels that refer to the same atoms, those of the inhomogeneities of the specimen.\\
On the other hand, one can write the second term of the second member of this (M.9), and therefore the second term of the second member of (M.8), as, since $P$ and $Q$ take the same values and $Q_{1}$ and $P_{1}$ take the same values:
\[
\sum_{P_{1}=1}^{M}\sum_{Q=1}^{N} J_{P_{1}Q}=\sum_{P=N-(M-1)}^{N}\sum_{Q=1}^{N} J_{PQ}=\sum_{Q_{1}=1}^{M}\sum_{P=1}^{N} J_{Q_{1}P}=\sum_{P=1}^{N} \sum_{Q_{1}=1}^{M}J_{Q_{1}P}
\]
and compare it with the third term of the second member of, also, (M.9) - and therefore with the third term of the second member of (M.8) - to observe that, provided $J_{Q_{1}P}=J_{PQ_{1}}$ or equivalently $J_{QP}=J_{PQ}$, i.e. the permutation of $P$ and $Q_{1}$ leaves $J_{PQ_{1}}$ unchanged or equivalently the permutation of $P$ and $Q$ leaves $J_{PQ}$ unchanged, the second and the third terms of the second member of (M.9) are equal to one another, i.e.
\begin{equation*}
\sum_{P_{1}=1}^{M}\sum_{Q=1}^{N} J_{P_{1}Q}= \sum_{P=1}^{N}\sum_{Q_{1}=1}^{M}J_{PQ_{1}}
\end{equation*}
which, in turn, allows writing (M.9) as:
\begin{equation*}
A= \sum_{P=1}^{N}\sum_{Q=1}^{N}J_{PQ} - 2 \sum_{P=1}^{N}\sum_{Q_{1}=1}^{M}J_{PQ_{1}}+\sum_{P_{1}=1}^{M}\sum_{Q_{1}=1}^{M}J_{P_{1}Q_{1}}  \tag{M.10}
\end{equation*}
where the first term of the second member represents the homogeneous material, i.e. the material in absence of inhomogeneities, the second term represents the coupling between the homogeneous material and the inhomogeneities and the third term represents the coupling between the inhomogeneities, the coupling between the local deviations from homogeneity. \\
We concludes from this that, provided the permutation of the labels in $J_{PQ}$ leaves $J_{PQ}$ unchanged, any equation of the type of (M.4) may be expressed as a sum of three components and be given by a relation of the sort of (M.10).\\
Thereafter, replacing $M$ by its expression (M.2) in the second term of (M.10) and successively applying the previous result, allows us to express this term as a sum of the type:
\begin{equation}
A_{2} =-2\sum_{P=1}^{N}\sum_{Q_{1}=1}^{M}J_{PQ_{1}}=-2\left[ \sum_{P=1}^{N}\sum_{i=1}^{n}\sum_{q_{1}^{\prime
}=1}^{M_{2i}}J_{Pq_{1}^{\prime }}-\sum_{P=1}^{N}\sum_{i=1}^{n}\sum_{q_{i}=1}^{M_{1i}}J_{Pq_{1}}\right] \text{.} \tag{M.11}
\end{equation}
A similar procedure leads to expressing the third term of (M.10) as:
\begin{equation}
A_{3} = \sum_{P_{1}=1}^{M}\sum_{Q_{1}=1}^{M}J_{P_{1}Q_{1}}=\sum_{i=1}^{n}%
\sum_{j=1}^{n}\sum_{p_{1}^{\prime }=1}^{M_{2i}}\sum_{q_{j}^{\prime
}=1}^{M_{2j}}J_{p_{i}^{\prime }q_{j}^{\prime }}
-2\sum_{i=1}^{n}\sum_{j=1}^{n}\sum_{p_{1}^{\prime
}=1}^{M_{2i}}\sum_{q_{j}=1}^{M_{1j}}J_{p_{i}^{\prime
}q_{j}}+\sum_{i=1}^{n}\sum_{j=1}^{n}\sum_{p_{1}=1}^{M_{1i}}%
\sum_{q_{j}=1}^{M_{1j}}J_{p_{i}q_{j}}\text{.}  \tag{M.12}
\end{equation}
Consequently and finally, provided the permutation of the labels in $J_{PQ}$ leaves $J_{PQ}$ unchanged, we may express (M.1) as follows:
\begin{equation}
A=\sum_{P=1}^{N}\sum_{Q=1}^{N}J_{PQ}+A_{2}\text{[equation(M.11)]}+A_{3}\text{%
[equation(M.12)].}  \tag{M.13}
\end{equation}
\rule[0mm]{\linewidth}{3.pt}
\section{Examples 1: diffraction intensities inferred}
\label{Examples1}
\subsection{Diffraction intensity from local deviations each of which possesses a center of symmetry}
\label{subDifLocDev}
At this level, the scattering from local deviations acting as if they were totally isolated from the host matrix, given by (\ref{8}), is expressed in terms of factors that are associated with the different, deficient and substitute, particles. The only assumption made here is that which was first made and used by Guinier, see \cite{Guinier (1939)}, in connection with centrosymmetrical particles. This assumption considers that each particle, here of either deficient or substitute type, possesses a center of symmetry. This is not a complex case.
\begin{figure}
\begin{center}
\includegraphics[scale=0.66]{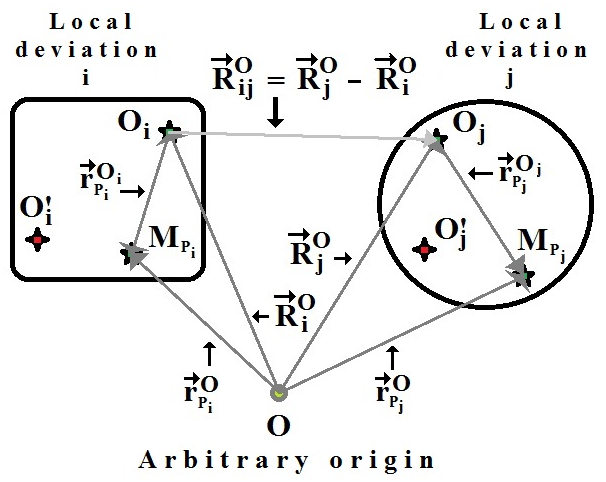}  
\textbf{\caption{Two, $i$ and $j$, local deviations from homogeneity.}}
\begin{minipage}[t]{14cm}
Two local deviations, $i$ and $j$, are shown. O$_{i}$ and O$_{i}^{\prime }$ are the centers of symmetry of, respectively, the substitute group $i$ and the deficient group $i$. O$_{j}$ and O$_{j}^{\prime }$ are similar to O$_{i}$ and O$_{i}^{\prime }$ but are relative to the local deviation $j$. $\vec{R}_{i}^{O} $ and $\vec{R} _{j}^{O}$ are the vectors giving the positions of the points O$_{i}$ and O$_{j}$ with respect to the arbitrary origin O; $\vec{R}_{i}^{\prime O}$ and $\vec{R}_{j}^{\prime O}$ locate the points O$_{i}^{\prime }$ and O$_{j}^{\prime }$, these are not represented in the figure.
\end{minipage}
\label{FigureInhom01}
\end{center}
\end{figure}
\noindent \underline{Substitute particle:} The center of symmetry for the $i^{th}$ substitute particle, the point $O_{i}$ in figure \ref{FigureInhom01}, may be defined as, since in a centrosymmetrical group of atoms an atom at a position given by $\vec{r}$ relative to the center of symmetry is coupled to another, identical, atom located at $-\vec{r}$ by the interatomic distance $\vert \vec{r} -(-\vec{r})\vert$ the middle point of which is the center of symmetry: 
\[
\sum f_{p_{i}}\overrightarrow{O_{i}M_{p_{i}}}=\sum f_{p_{i}}\vec{r}_{p_{i}}=0%
\text{,}
\]
where $\vec{r}_{p_{i}}=\overrightarrow{O_{i}M_{p_{i}}}$ is the vector which defines the position of the atom $p_{i}$ of the $i^{th}$ substitute particle with respect to $O_{i}$, its center of symmetry;\\
\underline{Deficient particle:} For the $i^{th}$ deficient particle the center of symmetry is defined by:
\[
\sum f_{p_{i}^{\prime }}\overrightarrow{O_{i}^{\prime }M_{p_{i}^{\prime }}}%
=\sum f_{p_{i}^{\prime }}\vec{r}_{p_{i}^{\prime }}=0\text{,} 
\]
where $\vec{r}_{p_{i}^{\prime }}=\overrightarrow{O_{i}^{\prime }M_{p_{i}^{\prime }}}$ gives the position of atom $p_{i}^{\prime }$ of the $ i^{th}$ deficient particle with respect to the center of symmetry $ O_{i}^{\prime }$. \\
We note that, according to figure 1, $O_{i}^{\prime }$ and $O_{i}$ do not necessarily coincide and $\vec{r}_{p_{i}q_{j}}$ may be written as
\begin{equation}
\vec{r}_{p_{i}q_{j}}=\vec{r}_{q_{j}}^{O_{j}}-\vec{r}_{p_{i}}^{O_{i}}+%
\vec{R}_{j}^{O}-\vec{R}_{i}^{O}=\vec{r}_{q_{j}}-\vec{r}_{p_{i}}+\vec{R}_{ij}%
\text{,}  \label{9}
\end{equation}
where 
\[
\vec{R}_{ij}=\vec{R}_{j}^{O}-\vec{R}_{i}^{O} 
\]
is the vector which gives the relative positions of the substitute particles $j$ and $i$, $O$ is the arbitrary origin. Similarly, one may write for the deficient particles:
\begin{equation}
\vec{r}_{p_{i}^{\prime }q_{j}^{\prime }}=\vec{r}_{q_{j}^{\prime }}-\vec{r}%
_{p_{i}^{\prime }}+\vec{R}_{ij}^{\prime }  \label{10}
\end{equation}
and for the deficient-substitute particle coupling:
\begin{equation}
\vec{r}_{p_{i}^{\prime }q_{j}}=\vec{r}_{q_{j}}-\vec{r}_{p_{i}^{\prime }}+(%
\vec{R}_{j}-\vec{R}_{i}^{\prime })  \label{11}
\end{equation}
where $\vec{R}_{j}-\vec{R}_{i}^{\prime }=\overrightarrow{R_{j}^{O}}-\overrightarrow{R_{i}^{\prime O}}$ gives the relative positions of the substitute ($j$) and the deficient ($i$) particles.\\
\textit{\textbf{Calculation of the terms of $I_{DD}(\vec{S})$ using }}(\ref{8}).
Inserting $\vec{r}_{p_{i}q_{j}}$, as given by (\ref{9}), into the cosine factor of the third term of (\ref{8}) gives:
\[
\cos (\vec{S}.\vec{r}_{p_{i}q_{j}})=\cos\vec{S}.(\vec{r}_{q_{j}}-\vec{r}_{p_{i}}+\vec{R}_{ij})
\]
and expanding gives:
\[
\cos (\vec{S}.\vec{r}_{p_{i}q_{j}})=\cos \vec{S}.(\vec{r}_{q_{j}}-\vec{r}%
_{p_{i}})\cos (\vec{S}.\vec{R}_{ij})-\sin \vec{S}.(\vec{r}_{q_{j}}-\vec{r}%
_{p_{i}})\sin (\vec{S}.\vec{R}_{ij})
\]
and, therefore: 
\begin{eqnarray}
\cos (\vec{S}.\vec{r}_{p_{i}q_{j}}) &=&[\cos \vec{S}.\vec{r}_{q_{j}}\cos 
\vec{S}.\vec{r}_{p_{i}}+\sin \vec{S}.\vec{r}_{q_{j}}\sin \vec{S}.\vec{r}%
_{p_{i}}]\cos \vec{S}.\vec{R}_{ij}  \nonumber \\
&&-[\sin \vec{S}.\vec{r}_{q_{j}}\cos \vec{S}.\vec{r}_{p_{i}}-\cos \vec{S}.%
\vec{r}_{q_{j}}\sin \vec{S}.\vec{r}_{p_{i}}]\sin \vec{S}.\vec{R}_{ij}\text{.}
\label{12}
\end{eqnarray}
So the third term of (\ref{8}) is obtained by inserting $\cos (\vec{S}.\vec{r}_{p_{i}q_{j}})$, as given by (\ref{12}), into this term and summing over all atoms $p_{i(j)}$ of all substitute particles $i(j)$. This contribution represents the scattering from the substitute particles only, i.e. the scattering in the ``absence of the host material''; writing it as $I_{DD3}(\vec{S})$ gives:
\begin{eqnarray}
I_{DD3}(\vec{S})
&=&\sum_{i=1}^{n}\sum_{j=1}^{n}\{(F_{1j}F_{1i}+F_{1j}^{\prime
}F_{1i}^{\prime })\cos \vec{S}.\vec{R}_{ij}  \nonumber \\
&&-(F_{1j}^{\prime }F_{1i}-F_{1j}F_{1i}^{\prime })\sin \vec{S}.\vec{R}%
_{ij}\},  \label{13}
\end{eqnarray}
where:
\[
F_{1j}=\sum_{q_{j}=1}^{M_{1j}}f_{q_{j}}\cos \vec{S}.\vec{r}_{q_{j}}\text{;
and }F_{1i}=\sum_{p_{i}=1}^{M_{1i}}f_{p_{i}}\cos \vec{S}.\vec{r}_{p_{i}} 
\]
are, as defined by Guinier, see \cite{Guinier (1939)} and \cite{Guinier and Fournet (1955)}, for the centrosymmetric particles, the structure factors of, respectively, the $j^{th}$ and the $i^{th}$ substitute particles. On the other hand, the factors $F_{1j}^{\prime }$ and $F_{1i}^{\prime }$ are, respectively, given by
\[
F_{1j}^{\prime }=\sum_{q_{j}=1}^{M_{1j}}f_{q_{j}}\sin \vec{S}.\vec{r}_{q_{j}}%
\text{, and }F_{1i}^{\prime }=\sum_{p_{i}=1}^{M_{1i}}f_{p_{i}}\sin \vec{S}.%
\vec{r}_{p_{i}}\text{ .} 
\]
Proceeding in the same way we find the first term, $I_{DD1}(\vec{S})$, of (\ref{8}) to be:
\begin{eqnarray}
I_{DD1}(\vec{S})
&=&\sum_{i=1}^{n}\sum_{j=1}^{n}\{(F_{2j}F_{2i}+F_{2j}^{\prime
}F_{2i}^{\prime })\cos \vec{S}.\vec{R}_{ij}  \nonumber \\
&&-(F_{2j}^{\prime }F_{2i}-F_{2j}F_{2i}^{\prime })\sin \vec{S}.\vec{R}_{ij}\}
\label{14}
\end{eqnarray}
where $F_{2j}$ and $F_{2i}$ are, respectively, the structure factors of the $j^{th}$ and the $i^{th}$ deficient centrosymmetric particles and are given by:
\[
F_{2j}=\sum_{q^{\prime}_{j}=1}^{M_{2j}}f_{q^{\prime}_{j}}\cos \vec{S}.\vec{r}_{q^{\prime}_{j}}\text{
and }F_{2i}=\sum_{p^{\prime}_{i}=1}^{M_{2i}}f_{p^{\prime}_{i}}\cos \vec{S}.\vec{r}_{p^{\prime}_{i}}%
\text{.} 
\]
The $F_{2j}^{\prime }$ and $F_{2i}^{\prime }$ factors are as follows:
\[
F_{2j}^{\prime }=\sum_{q^{\prime}_{j}=1}^{M_{2j}}f_{q^{\prime}_{j}}\sin \vec{S}.\vec{r}_{q^{\prime}_{j}}%
\text{ and }F_{2i}^{\prime }=\sum_{p^{\prime}_{i}=1}^{M_{2i}}f_{p^{\prime}_{i}}\sin \vec{S}.%
\vec{r}_{p^{\prime}_{i}}\text{.} 
\]
Equation (\ref{14}) represents the scattering from the deficient particles alone.\\
Finally, we find the second term of (\ref{8}) is given by: 
\begin{eqnarray}
I_{DD2}(\vec{S})
&=&-2n\sum_{i=1}^{n}\sum_{j=1}^{n}\{(F_{1j}F_{2i}+F_{1j}^{\prime
}F_{2i}^{\prime })\cos \vec{S}.(\vec{R}_{j}-\vec{R}_{i}^{\prime })  \nonumber
\\
&&-(F_{1j}^{\prime }F_{2i}-F_{1j}F_{2i}^{\prime })\sin \vec{S}.(\vec{R}_{j}-%
\vec{R}_{i}^{\prime })\}  \label{15}
\end{eqnarray}
This term, $I_{DD2}(\vec{S})$, represents the coupling of the deficient and the substitute particles. $F_{1j}$, $F_{2i}$, $F_{1j}^{\prime }$ and $F_{2i}^{\prime }$ are as defined above and together with the other $F_{2j}$, $F_{2j}^{\prime }$ etc. will be referred to, in the remaining part of this text, as the ``$F$'' factors.\\
The scattered contribution $I_{DD}(\vec{S})$, see (\ref{8}), to the total scattered intensity from the ensemble of the local deviations ``behaving as if they were totally isolated from the host matrix'' is obtained by combining equations (\ref{13}), (\ref{14}) and (\ref{15}).\\
No structure-related assumption has been made so far apart from allowing ($\alpha $) the different atomic structure factors to be real and ($\beta $) all the particles to have a center of symmetry. Therefore, equations (\ref{13}), (\ref{14}) and (\ref{15}) are expected to hold in any structural situation, that satisfies ($\alpha $) and ($\beta $), for which the relaxation effects that may result from the particle substitutions are negligible. In particular, they should hold in the case of non-interacting as well as in the case of interacting local deviations. A limited treatment of these two cases is given in the next subsections.
\subsection{Scattering from \textit{n} non-interacting local deviations}
\subsubsection{The not necessarily identical deviations}
Widely separated deviations could still have some interacting effects. However, these are commonly assumed to be negligible and ignored. The main consequence of this approximation is to ignore, in the expression of the
scattered intensity, all terms that are not due to the self-pairing of the different local deviations. This is achieved by setting $i=j$ in all equations that describe the scattered intensity. Consequently:
\[
\vec{R}_{ij}=\vec{R}_{ij}^{\prime }=\vec{R}_{ii}=\vec{R}_{ii}^{\prime }=\vec{%
0}\text{ and }\vec{R}_{j}-\vec{R}_{i}^{\prime }=\vec{R}_{i}-\vec{R}_{i}=%
\overrightarrow{O^{\prime }O_{i}}\text{,} 
\]
where $\overrightarrow{O^{\prime }O_{i}}$ is the vector which gives the relative positions of the centers of symmetry of the substitute and the deficient particles of the $i^{th}$ local deviation, therefore:
\[
F_{1j}=F_{1i}\text{; }F_{1j}^{\prime }=F_{1i}^{\prime }\text{; }F_{2j}=F_{2i}%
\text{; }F_{2j}^{\prime }=F_{2i}^{\prime }\text{; }M_{1j}=M_{2i}\text{ and }%
M_{2j}=M_{2i}\text{.} 
\]
Using these results together with equations (\ref{13}), (\ref{14}) and (\ref{15}) leads to the following expression for the contribution from non-interacting local deviations, $I_{DD}(\vec{S})$:
\begin{eqnarray}
I_{DD}(S) &=&\sum_{i=1}^{n}\left\lbrace F_{1i}^{2}+F_{1i}^{\prime
2}+F_{2i}^{2}+F_{2i}^{\prime 2}+2[F_{1i}F_{2i}+F_{1i}^{\prime
}F_{2i}^{\prime }]\cos \vec{S}.\overrightarrow{O^{\prime }O_{i}}\right.  \nonumber
\\
&&\left.-2[F_{1i}^{\prime }F_{2i}-F_{1i}F_{2i}^{\prime }]\sin \vec{S}.%
\overrightarrow{O^{\prime }O_{i}}\right\rbrace \text{.}  \label{16}
\end{eqnarray}
The form of this relation suggests that it may be written as a sum of $n$ terms, $I_{iDD}(\vec{S})$, $i = 1, 2, \cdots, n$, each, $i$, of which represents the contribution from a single local deviation: 
\[
I_{DD}(\vec{S})=\sum_{i=1}^{n}I_{iDD}(\vec{S})\text{.} 
\]
\subsubsection{The non-interacting and identical local deviations}
In the case where all local deviations are identical, or assumed to be, on average, identical, one may expect, as a consequence, that all deficient particles are similar and that all substitute particles are similar. This would imply that particles of the same type consist of atoms of the same chemical natures which are distributed according to a same atomic distribution function, therefore the number of atoms $M_{2i}$ is expected to be the same for all deficient particles ($M_{2i}=M_{2}$ whatever $i$) and, similarly, $M_{2i}=M_{2}$ whatever $i$ for the substitute particles. Although these considerations leave the expression for the scattered component $I_{DD}(\vec{S})$, see (\ref{16}), formally unchanged, only $M_{1i}$ and $M_{2i}$ are set equal to, respectively, $M_{1}$ and $M_{2}$, they greatly ease the calculations for two limiting cases: the case of identical and randomly oriented local deviations and the case of identical local deviations with a preferred orientation. These two particular cases are considered next.
\paragraph{The case of identical and randomly oriented local deviations} 
The first term of $I_{DD}(\vec{S})$, (\ref{8}), corresponds to the contribution from the substitute particles ``behaving as if they were isolated from the host matrix''. The third term, on the other hand, corresponds to the contribution from the deficient particles acting as if they were alone. These two terms are similar to one another and to the expression for the relevant corresponding system, or ensemble, of groups of atoms, or particles, considered and dealt with elsewhere by Guinier, see \cite{Guinier (1939)} and \cite{Guinier and Fournet (1955)}. But the second term of this equation is the result of the coupling between the deficient particles and the substitute particles. This is similar in form to the scattering from a set of simple particles, the difference is that $\vec{r}_{p^{\prime }q}$ does not necessarily take zero values.\footnote{The exception is for the case where the atomic distribution functions for both deficient and substitute particles are the same.} For randomly oriented identical deviations, we find, through angularly averaging this second term over the $4\pi$ solid angle as done by Debye to get the equation bearing his name, i.e. the Debye equation of scattering also known as the Debye equation, see \cite{Debye 1915}, \cite{Guinier (1939)}, \cite{Guinier and Fournet (1955)}, that $I_{DD}(\vec{S})$ is given by:
\begin{eqnarray}
I_{DD}(S) &=&n\left\lbrace \sum_{p=1}^{M_{1}}\sum_{q=1}^{M_{1}}f_{p}f_{q}\frac{%
\sin Sr_{pq}}{Sr_{pq}}-2\sum_{p^{\prime
}=1}^{M_{2}}\sum_{q=1}^{M_{1}}f_{p^{\prime }}f_{q}\frac{\sin Sr_{p^{\prime
}q}}{Sr_{p^{\prime }q}} \right.  \nonumber \\
&&\left.+\sum_{p^{\prime }=1}^{M_{2}}\sum_{q^{\prime }=1}^{M_{2}}f_{p^{\prime
}}f_{q^{\prime }}\frac{\sin Sr_{p^{\prime }q^{\prime }}}{Sr_{p^{\prime
}q^{\prime }}}\right\rbrace   \label{17}
\end{eqnarray}
where $p^{\prime }$ and $q^{\prime }$ are the labels of the atoms of one of the deficient particles, and $p$ and $q$ are the labels of the atoms of one of the substitute particles. For monatomic systems, $I_{DD}(\vec{S})$ is obtained by setting $f_{p^{\prime }}=f_{q^{\prime }}$, whatever $p^{\prime }$ and $q^{\prime }$, and $f_{p}=f_{q}$, whatever $p$ and $q$.
\paragraph{Case of identical local deviations that are oriented in a same direction}
In this case the scattered component $I_{DD}(\vec{S})$ is found to be given by:
\begin{eqnarray}
I_{DD}(\vec{S}) &=&n\left\lbrace  \sum_{p=1}^{M_{1}}\sum_{q=1}^{M_{1}}f_{p}f_{q}\cos 
\vec{S}.\vec{r}_{pq}-2\sum_{p^{\prime
}=1}^{M_{2}}\sum_{q=1}^{M_{1}}f_{p^{\prime }}f_{q}\cos \vec{S}.\vec{r}%
_{p^{\prime }q}\right.  \nonumber \\
&&\left. +\sum_{p^{\prime }=1}^{M_{2}}\sum_{q^{\prime }=1}^{M_{2}}f_{p^{\prime
}}f_{q^{\prime }}\cos \vec{S}.\vec{r}_{p^{\prime }q^{\prime }}\right\rbrace   \label{18}
\end{eqnarray}
\section{Examples 2: the simplest ``Small Angle Scattering'' cases}
\label{Examples2}
SAS has found application in both material science and biology. Effectively, small-angle X-ray scattering (SAXS), sensitive to inhomogeneities in electron density in the 10-250-nm range, is very useful for the study of materials such as the organic-templated mesostructured chalcogenides, \cite{Brant et al (2013)}, and, together with small-angle neutron scattering (SANS), is useful for investigating the nanoscale structure and interactions in materials with disorder at the atomic scale level, see \cite{Eds Duncan et al 2014} plus references therein. Also, SAXS is, at present and for various reasons, the standard tool for biologists, see \cite{Grawert and Svergun 2020}. 
In this section we consider two SAS cases. One is relative to identical non-interacting and randomly oriented local deviations, see  \ref{ROIandN-ILD}, and the other one is relative to identical non-interacting with ``a preferred orientation'' local deviations, see \ref{IN-IwPOLD}.
The scattering which appears at small angles, around the direction of the unscattered beam, may be dominated by the scattering from finite sized particles embedded in an otherwise homogeneous material, see \cite{Guinier (1939)}, \cite{Guinier and Fournet (1955)}. We, therefore, may expect the small angle scattering from a real specimen to be essentially coming from the ensemble of its local deviations and given by the scattering component which is represented by one of the two (\ref{17}) and (\ref{18}) according to whether the deviations are randomly oriented or oriented in a well defined direction. The two terms corresponding, one, to the deficient and, the other one, to the substitute particles, i.e. the first and the third terms of either of (\ref{17}) and (\ref{18}), acting as if they were alone are similar to the original cases considered by Guinier, \cite{Guinier (1939)}, and therefore have been dealt with elsewhere, -see also \cite{Guinier and Fournet (1955)}-, and were found to approximate a Gaussian form at small angles in the case of randomly oriented non-interacting particles. The approximate expression for the second term of (\ref{17}) is obtained here by applying the technique by Guinier, see \cite{Guinier (1939)}, and found to be, see next subsection, a bit different from the Guinier law: it contains in it an extra local deviation-related parameter.
\subsection{Randomly oriented identical and non-interacting local deviations}
\label{ROIandN-ILD}
Applying the method by Guinier for randomly oriented identical non-interacting local deviations, the deficient-substitute coupling term, $I_{DD2}(S)$ as given by the second term of (\ref{17}), is expanded in a power series of $Sr_{p^{\prime }q}$ to give, when $Sr_{p^{\prime }q}$ $<<1$:
\begin{eqnarray}
I_{DD2}(S) &\thickapprox &-2n\left[ \sum_{p^{\prime
}=1}^{M_{2}}\sum_{q=1}^{M_{1}}f_{p^{\prime }}f_{q}-\sum_{p^{\prime
}=1}^{M_{2}}\sum_{q=1}^{M_{1}}f_{p^{\prime }}f_{q}\frac{S^{2}}{6}%
r_{p^{\prime }q}^{2}\right.  \nonumber \\
&&\left.+\sum_{p^{\prime }=1}^{M_{2}}\sum_{q=1}^{M_{1}}f_{p^{\prime }}f_{q}\frac{%
S^{4}}{120}r_{p^{\prime }q}^{4}- \cdots \right]   \label{19}
\end{eqnarray}
where $r_{p^{\prime }q}=\mid \vec{r}_{q}-\vec{r}_{p}^{\prime }+\overrightarrow{O^{\prime }O\mid }$, which inserted into this limited expansion of $I_{DD2}(S)$, leads to a sum of a quite large number of terms. However, consequent to the allowance for a center of symmetry to exist for either of the deficient and substitute particles, this number reduces to ten, since each vector $\vec{r}$, e.g. $\vec{r}_{p}$, is dual to another vector $-\vec{r}$, e.g. $-r_{p}$, thus causing all terms involving odd powers of $\vec{r}$ to vanish. For instance the term which is related to $ \vec{r}_{q}.(\vec{r}_{p^{\prime }})^{3}$ is given by:
\begin{equation}
8n\frac{S^{4}}{120}\sum_{q=1}^{M_{1}}\sum_{p^{\prime
}=1}^{M_{2}}f_{q}f_{p^{\prime }}\vec{r}_{q}(\vec{r}_{p^{\prime }})^{3}=8n\frac{%
S^{4}}{120}\left( \sum_{q=1}^{M_{1}}f_{q}\vec{r}_{q}\right) \left(
\sum_{p^{\prime }=1}^{M_{2}}f_{p^{\prime }}(\vec{r}_{p^{\prime }})^{3}\right)
=0\text{,}  \label{20}
\end{equation}
where each of the sums in brackets is zero. And thus $I_{DD2}(S)$ reduces to:
\begin{eqnarray}
I_{DD2}(S) &\thickapprox &-2n\left( \sum_{p^{\prime }=1}^{M_{2}}f_{p^{\prime
}}\right) \left( \sum_{q=1}^{M_{1}}f_{q}\right) \{1-\frac{S^{2}}{6}%
(R^{2}+R^{\prime 2}+O^{\prime }O^{2})  \nonumber \\
&&+\frac{S^{4}}{120}[R^{4}+R^{\prime 4}+O^{\prime }O^{4}+6(R^{2}R^{\prime
2}+R^{2}O^{\prime }O^{2}  \nonumber \\
&&\qquad \qquad \quad +R^{\prime 2}O^{\prime }O^{2})]-...\}\text{,}
\label{21}
\end{eqnarray}
[i.e. after dividing and multiplying by $(\sum_{p^{\prime }=1}^{M_{2}}f_{p^{\prime }})(\sum_{q=1}^{M_{1}}f_{q})$] where:
\begin{equation}
R^{\prime 2}=\frac{\sum_{p^{\prime }=1}^{M_{2}}f_{p^{\prime }}\vec{r}%
_{p^{\prime }}^{2}}{\sum_{p^{\prime }=1}^{M_{2}}f_{p^{\prime }}}\text{ \quad
and\quad }R^{2}=\frac{\sum_{q=1}^{M_{1}}f_{q}\vec{r}_{q}^{2}}{%
\sum_{q=1}^{M_{1}}f_{q}}\text{,}  \label{22}
\end{equation}
$R$ and $R^{\prime }$ are the radii of gyration, as defined by Guinier, see \cite{Guinier (1939)} and \cite{Guinier and Fournet (1955)}. These are associated with, respectively, the substitute and the deficient particles and are dimensions dependent structural parameters. Thus, doing as Guinier did, \cite {Guinier (1939)}, \cite{Guinier and Fournet (1955)}, and provided the contribution from the terms in $S^{4}$ and higher are not significant we may write (\ref{19}), at small angles, as;
\begin{equation}
I_{DD2}(S)\thickapprox -2n\left( \sum_{p^{\prime }=1}^{M_{2}}f_{p^{\prime
}}\right) \left( \sum_{q=1}^{M_{1}}f_{q}\right) \exp \left( -\frac{%
R^{2}+R^{\prime 2}+O^{\prime }O^{2}}{6}S^{2}\right)  \label{23}
\end{equation}
where $O^{\prime }O$ is the distance between the centers of symmetry of the deficient and the substitute particles that pertain to the same local deviation. It is to be stressed, at this point, that this expression is only valid over the angular range where:
\[
(R^{2}+R^{\prime 2}+O^{\prime }O^{2})S^{2}/6<<1\text{,}
\]
for then it makes a good approximation to the exact expression, i.e. the second term of (\ref{17}). Thus, the total small angle scattering from $n$ identical local deviations that are randomly oriented is approximately given by:
\begin{eqnarray}
I_{SAS}(S) &\thickapprox &n\left( \sum_{p=1}^{M_{1}}f_{p}\right) ^{2}\exp
\left( -\frac{S^{2}R^{2}}{3}\right) +n\left( \sum_{p^{\prime
}=1}^{M_{2}}f_{p^{\prime }}\right) ^{2}\exp \left( -\frac{S^{2}R^{\prime 2}}{%
3}\right)  \nonumber \\
&&-2n\left( \sum_{p=1}^{M_{1}}f_{p}\right) \left( \sum_{p^{\prime
}=1}^{M_{2}}f_{p^{\prime }}\right) \exp \left( -\frac{S^{2}(R^{2}+R^{\prime
2}+O^{\prime }O^{2}}{6}\right) \text{,}  \label{24}
\end{eqnarray}
where:
\[
\frac{S^{2}R^{\prime 2}}{3}<<1\text{; }\frac{S^{2}R^{2}}{3}<<1\text{ and }%
\frac{S^{2}(R^{2}+R^{\prime 2}+O^{\prime }O^{2})}{6}<<1\text{.} 
\]
The shape of (\ref{24}) is different from the Gaussian shape. Therefore the application of the Guinier law is not possible in this case.\\
Also, it is important to notice that (\ref{24}) is an \emph{approximate} expression for the corresponding exact expression, given by (\ref{17}), which is suitable for use when and only when the conditions imposed by $S^{2}R^{\prime 2}/3<<1\text{; }S^{2}R^{2}/3<<1\text{ and }S^{2}(R^{2}+R^{\prime 2}+O^{\prime }O^{2})/6<<1$ are fully satisfied. These conditions totally specify the angular range of validity of (\ref{24}); and within this range (\ref{24}) varies rather like $y=a-bS^{2}+\cdots$. Consequently, if (\ref{24}) is applied properly, i.e. really within the angular range over which it is effectively valid, then the exponential decays it is made up of will introduce no apparent correlation peak within this range.\\
If furthermore the host matrix and the substitute particles are monatomic then, in this equation, $(\sum_{p=1}^{M_{1}}f_{p})^{2}$, $(\sum_{p^{\prime}=1}^{M_{2}}f_{p^{\prime }})^{2}$ and $(\sum_{p=1}^{M_{1}}f_{p})(\sum_{p^{\prime }=1}^{M_{2}}f_{p^{\prime }})$ become, respectively, $(M_{1}f_{B})^{2}$, $(M_{2}f_{M})^{2}$ and $M_{1}M_{2}f_{M}f_{B}$, where $f_{B}$ and $f_{M}$ are the atomic scattering factors for, respectively, the substitute and the host matrix atoms.\\
We note that $OO^{\prime }=O^{\prime }O$ is expected to be different from zero in situations where the structural differences, in size and in shape, and therefore in dimensions, between the deficient and the substitute particles are such that their respective centers of symmetry do not coincide. For instance, this can be so when the size of the substitute atoms is comparatively large or comparatively small with respect to the size of the deficient atoms.\\
This subsection on small angle scattering is closed with a table of some particular expression cases deduced from (\ref{24}). Therefore, all these expression cases deduce from the mother expression for the partial intensity, $I_{DD}(\vec{S})$, given by (\ref{8}) and all of them rely on the approximations used by Guinier to get the law bearing his name, the Ginier law. These approximations, in number of two, are: that assuming the randomness considered by Debye, \cite{Debye 1915}, to get the formula bearing his name, i.e. the Debye formula of scattering, and that assuming the product of $S$ and $r$ is very small compared with unity considered by Guinier, \cite{Guinier (1939)}, e.g. $Sr_{p^{\prime }q}$ $<<1$. Therefore the application of all of these expressions are subject to the same sort of restrictions as the Guinier law. The approximation considered by Guinier is particularly restricting as regards to the angular range over which his law applies. 
\subsubsection{A table of SAS intensity expressions for particular inhomogeneity cases obtained from (\ref{24})}
\paragraph{Case 1}
Case where the centers of symmetry of the deficient and the substitute particles coincide, i.e. $OO^{\prime }=0$;
\begin{equation}
I_{SAS}(S)\thickapprox n\left\{ \left[ \sum_{p=1}^{M_{1}}f_{p}\right] \exp
\left( -\frac{S^{2}R^{2}}{6}\right) -\left[ \sum_{p^{\prime
}=1}^{M_{2}}f_{p^{\prime }}\right] \exp \left( -\frac{S^{2}R^{\prime 2}}{6}%
\right) \right\} ^{2}\text{.}  \label{25}
\end{equation}
This (\ref{25}) resembles, but is fundamentally different from, the equation given at the bottom of page 342 in the Guinier's 1963 book, \cite{Guinier (1963)}, which is quite normal, since the two equations correspond to two different material systems.
\paragraph{Case 2}
Case where the atomic distributions of the deficient and the substitute particles are identical, but with different atomic scattering factors. In this situation we may expect the number of atoms to be the same ($M$) for both types of particles. Therefore, we may expect the radii of gyration to be the same ($R=R^{\prime }$) for both types of particles and write
\begin{equation}
I_{SAS}(S)\thickapprox n\left\{ \left[ \sum_{p=1}^{M}f_{p}\right] -\left[
\sum_{p^{\prime }=1}^{M}f_{p^{\prime }}\right] \right\} ^{2}\exp \left( -%
\frac{S^{2}R^{2}}{3}\right) \text{.}  \label{26}
\end{equation}
If, furthermore, in this particular case, we assume the scattering is associated with some average scattering power densities 
\[
\rho _{s}=\frac{\sum_{p=1}^{M}f_{p}}{V}
\]
for the substitute particles and 
\[
\rho _{d}=\frac{\sum_{p^{\prime }=1}^{M}f_{p^{\prime }}}{V}
\]
for the deficient particles; where $V$ is the volume that is occupied by one local deviation, then (\ref{26}) becomes
\begin{equation}
I_{SAS}(S)\thickapprox nV^{2}[\rho _{s}-\rho _{d}]^{2}\exp \left( -\frac{%
S^{2}R^{2}}{3}\right) \text{.}  \label{27}
\end{equation}
\paragraph{Case 3}
In situations where the local deviations are associated with pure deficiencies, e.g. when dealing with pure materials, the small angle scattering approximate distribution is obtained by setting equal to zero the
scattering factors of the substitute atoms in (\ref{24}) to get to the form expressing the very well known Guinier law, see \cite {Guinier (1939)}, \cite{Guinier and Fournet (1955)}, \cite {Guinier (1963)}:
\begin{equation}
I_{SAS}(S)\thickapprox n\left[ \sum_{p^{\prime }=1}^{M_{2}}f_{p^{\prime
}}\right] ^{2}\exp \left( -\frac{S^{2}R^{\prime 2}}{3}\right) \text{,}
\label{28}
\end{equation}
$R^{\prime }$ is the radius of gyration for the deficient particles.
\subsection{Identical non-interacting with ``a preferred orientation'' local deviations}
\label{IN-IwPOLD}
For identical non-interacting and ``all oriented in the same direction'' local deviations the scattered intensity component $I_{DD}(S)$, as given by (\ref{18}), may be approximated using the method given in the book by Guinier and Fournet, see \cite{Guinier and Fournet (1955)}, to give:
\begin{eqnarray}
I_{SAS}(S) &\thickapprox
&n\sum_{p=1}^{M_{1}}\sum_{q=1}^{M_{1}}f_{p}f_{q}\exp [-S^{2}D^{2}(\vec{u} )]+n\sum_{p^{\prime }=1}^{M_{2}}\sum_{q^{\prime }=1}^{M_{2}}f_{p^{\prime }}f_{q^{\prime }}\exp [-S^{2}D^{\prime 2}(\vec{u})]  \nonumber \\
&&-2n\sum_{p^{\prime }=1}^{M_{2}}\sum_{q=1}^{M_{1}}f_{p}f_{q}\exp \left[ -\frac{S^{2}(D^{2}(\vec{u})+D^{\prime 2}(\vec{u})+(\vec{u}.OO^{\prime })^{2})}{2}\right] \text{.}  \label{29}
\end{eqnarray}
Where: 
\begin{equation}
D^{2}(\vec{u})=\frac{\sum_{q=1}^{M_{1}}f_{q}(\vec{u}.\vec{r}_{q})^{2}}{%
\sum\nolimits_{q=1}^{M_{1}}f_{q}}\text{\quad and\quad }D^{\prime 2}(\vec{u})=%
\frac{\sum\nolimits_{p^{\prime }=1}^{M_{2}}f_{p^{\prime }}(\vec{u}.\vec{r}%
_{p^{\prime }q})^{2}}{\sum\nolimits_{p^{\prime }=1}^{M2}f_{p^{\prime }}}%
\text{.}  \label{30}
\end{equation}
$D^{2}(\vec{u})$ is the square of the average inertial distance $D(\vec{u})$, see, for instance, the book by Guinier and Fournet \cite{Guinier and Fournet (1955)}, of the substitute particle with respect to the plane perpendicular to $\vec{u}$, the unit vector which defines the direction of the scattering vector $\vec{S}$, that passes through the center of symmetry $O$ of the substitute particle. $D^{\prime }(\vec{u})$ is the square of the average inertial distance $D^{\prime 2}(\vec{u})$ associated with the deficient particle with respect to the plane perpendicular to $\vec{u}$ and which passes through the center of symmetry $O^{\prime }$.
\section{Example 3: identical centrosymmetric interacting local deviations}
\label{Examples3}
In this section we deduce from (\ref{8}), as another example of application, the partial diffraction intensity expression, given by (\ref{34}), relative to identical centrosymmetric interacting local deviations which was originally obtained by Guinier and Fournet, see \cite{Guinier and Fournet (1955)}. The deduction requires passing through different successive intensity expressions each of which is, therefore, relative to a different stage corresponding to a different material case, the complexity of the material decreasing progressively.  

The total scattered intensity from an inhomogeneous system that may be described as being made up of $n$ identical interacting local deviations may be obtained by firstly adding together ($\delta $) the intensity contributions originating from the coupling of the local deviations with the ``homogeneous system'', ($\epsilon $) the intensity contributions that are due to the local deviations in the absence of the host matrix and ($\varepsilon $) the scattered intensity from the homogeneous system, i.e. in absence of the local disturbances, and secondly setting (i) $M_{1i(j)}$, the number of atoms which make up the $i(j)^{th}$ substitute particle, equal to the same number $M_{1}$ for all substitute particles and (ii) $M_{2i(j)}$ equal to the same number $M_{2}$ for all deficient particles. This leaves the total scattered intensity expression formally unchanged. However, it is considerably simplified in the particular case of monatomic systems, i.e. when the inclusion particles as well as the host matrix are monatomic. 

The assumption, see \cite{Guinier (1939)}, \cite{Guinier and Fournet (1955)} that the particles possess a center of symmetry allows writing: $F_{2i}^{\prime }=F_{2j}^{\prime }=F_{1j}^{\prime }=F_{1i}^{\prime }=0$, since, in this situation, each vector $\vec{r}$ of any particle possesses a dual vector $-\vec{r}$ with respect to the center of symmetry resulting in the zeroing of the primed $F$ factors. This leads, in a first stage, to writing the contribution to the total scattered intensity that is due to $n$ local interacting deviations, this is given by the sum of the three (\ref{13}), (\ref{14}) and (\ref{15}), as:
\begin{eqnarray}
I_{DD}(\vec{S}) &=&\sum_{i=1}^{n}\sum_{j=1}^{n}\left\lbrace [F_{1j}F_{1i}]\cos \vec{S}.\vec{R}_{ij}+[F_{2j}F_{2i}]\cos \vec{S}.\vec{R}_{ij}^{\prime }\right. \nonumber \\
&&\left.-2[F_{1j}F_{2i}]\cos \vec{S}.(R_{j}-\vec{R}_{i}^{\prime })\right\rbrace \text{,}
\label{31}
\end{eqnarray}
where the different $F_{1i(j)}$ and $F_{2i(j)}$ are structure factors. Further simplifications may be obtained for the scattering from identical local deviations if the centers of symmetry of the substitute and the deficient particles coincide. In this case, and a second stage, $\vec{R}_{j}=\vec{R}_{j}^{\prime }$ whatever $j$, and (\ref{31}) simplifies to lead to an $I_{DD}(\vec{S})$ given by:
\begin{equation}
I_{DD}(\vec{S})=\sum_{i=1}^{n}%
\sum_{j=1}^{n}[F_{1j}F_{1i}-2F_{1j}F_{2i}+F_{2j}F_{2i}]\cos \vec{S}.\vec{R}%
_{ij}\text{.}  \label{32}
\end{equation}
For pure materials, i.e. when no particle is included, therefore when the local deviations are of pure deficiency type, the scattering component $I_{DD}(\vec{S})$ is obtained by setting to zero the scattering factors of the substitute particles, $F_{1i(j)}$, in (\ref{32}) which, then and in a third and last stage, leads to an $I_{DD}(\vec{S})$ given by:
\begin{equation}
I_{DD}(\vec{S})=\sum_{i=1}^{n}\sum_{j=1}^{n}[F_{2j}F_{2i}]\cos \vec{S}.\vec{R%
}_{ij}  \label{33}
\end{equation}
and has the same form as (\ref{1}). This similarity suggests that, as for atoms, diffraction peaks may be generated by particles present in the material system, i.e. in favorable cases. The $F$ factors play here a role that is very much similar to that of the atomic form factors $f$ in (\ref{1}) and are particle shape and size, and therefore dimensions, dependent.
Finally, using the expressions for $F_{2i(j)}$, see \ref{subDifLocDev}, allows getting:
\begin{equation}
I_{DD}(\vec{S})=\sum_{i=1}^{n}\sum_{j=1}^{n}\left[ \sum_{p_{j}^{\prime }=1}^{M_{2}}f_{p_{j}^{\prime }}\cos \vec{S}.\vec{r}_{p_{j}^{\prime }}\right] \left[ \sum_{q_{j}^{\prime }=1}^{M_{2}}f_{q_{j}^{\prime }}\cos \vec{S}.\vec{r}_{q_{j}^{\prime }}\right] \cos \vec{S}.\vec{R}_{ij}\text{,}  \label{34}
\end{equation}
which is, as it should be, identical to the corresponding expression that is given in the book by Guinier and Fournet, \cite{Guinier and Fournet (1955)}, and that is relative to ``a group of identical interacting particles''.
\section{ Conclusion}
It has been found that the introduction of structural disturbances on an otherwise homogeneous material system yields a total scattered intensity distribution that is the sum of two main components. The first of these is that which one would expect if the material were in its homogeneous state, and the other, an additional contribution due to the ``particle substitution operation'', represents the deviation from homogeneity of the material system. The introduction of the structural disturbances, through substitutions of groups of atoms (small particles) by other groups of atoms of a different kind, allows varying at will the two important degrees of freedom that are the chemical nature and the atomic distribution. The additional scattering is found to consist of two separate groups of components. One of these represents the scattering from the local deviations acting as if they were totally isolated from the host matrix and the other represents the coupling between the local deviations and the ``host homogeneous state''. The presence of this last component indicates that the scattering from inhomogeneous systems may not be assumed to consist of the simple sum of the intensity from the ideal network, on the one hand, and of that from the inhomogeneities supposed totally isolated from the host material, on the other hand. Thus, this conceptional representation of a real material has resulted in a formulation of the scattered intensity that may prove to be useful for comparative studies of material systems with the same, or, at least, on average the same, sort of network. This is for the reason that the parts of the scattered intensity distribution which are due to the particles present within the material system are very much dependent on their dimensions.
It has also resulted in shapes of the scattered intensity distribution at small angles, i.e. from inhomogeneous material systems, that do not obey the very attractive Guinier law in some cases. As a consequence of this, the techniques of analysis [such as the Guinier plot \cite{Guinier (1939)},
\cite{Guinier and Fournet (1955)}, \cite {Guinier (1963)}, the tangent technique, \cite{Guinier and Fournet (1955)}, \cite{Guinier (1963)}, and the derivative of the logarithm of the small angle scattering intensity distribution technique, \cite{Hadji and Craven (1988)}], which can be used to extract the radius of gyration, first defined in \cite{Guinier (1939)}, are not applicable in cases where the chemical nature and the atom distribution of the inclusion particles do not allow representing the intensity at small angles of scattering using a Gaussian form. New techniques will, therefore, be most welcome.

\end{document}